# Temporal Retention of Information as a Biosignature

Terence P. Kee[1] and James McCrum[2]

[1] School of Chemistry, University of Leeds, Leeds, UK

[2] School of Physics, Engineering, and Technology, University of York, York, UK

**Corresponding author**: Terence P. Kee; Email: T.P.Kee@leeds.ac.uk

## Abstract

Previous publications by the authors put forward the argument that Lifelike Cellular Automata (LCAs) can be treated as a *bona fide* example of livingness in and of themselves, not simply a toy analogue to biological life. Traits known to be indicative of biological life – biosignatures – were identified in informational form as particular outlier traits of the ruleset for the LCA known as Conway's Game of Life (CGOL). This publication reverses that logic, looking at a known outlier trait of CGOL – its very long-lasting evolutions – and using this to point towards temporal retention as an informational biosignature concept.

## Introduction

Defining life is an infamously difficult task. This is an issue for astrobiologists, because in order to be able to identify and categorize life in the universe, first being able to *define* what it is that you're talking about is something of a necessary prerequisite. (Dunér, 2019; Gillen, 2023; Cowie, 2023).

To get an idea of how troublesome of a problem finding one definitive definition of life is, Edward Trifonov was able to find *one hundred and twenty three* different ways to do so. (Trifonov, 2011). He took a somewhat unorthodox approach to the problem from there, using a linguistic analysis to synthesize these many definitions down to simply:

*"Life is self-reproduction with variations."*

In practice, as this approach emphasizes, many definitions of life are strongly conceptually overlapping, often unintentionally. (Benner, 2010; Letelier, 2011; Cornish-Bowden, 2015; Cornish-Bowden, 2020). Many of these attempts have looked at life through the lens of information. (Walker, 2017; Helman, 2022; Miller, 2023; Bartlett, 2025).

This has been a point of some criticism; for example, Assembly Theory, an attempt to empirically quantify a concept of life, (Sharma, 2023), has been noted to be reducible to Shannon's Information Theory, which has been used to totally dismiss said definition. (Abrahao, 2024; Ozelim, 2024; Uthamacumaran, 2024).

As the authors noted in a previous publication (McCrum & Kee, 2024) the approach of turning to the general concept of information as an attempt to resolve the question of defining life can be seen as analogous to parallel trends in neuroscience. The model of Integrated Information Theory is a high-profile attempt to empirically quantify a concept of consciousness through the parameter of integrated information Φ (Tononi, 2004; Albantakis, 2023), although this has faced savage criticism on a variety of conceptual and structural fronts. (Barrett, 2019; Nizami, 2019; Doerig, 2019; Merker, 2021; Herzog, 2022; Fleming, 2023; Klincewicz, 2025).

In said previous publication, drawing on arguments in another publication in the same issue by the authors (Kee & McCrum, 2024), the authors made the case for a broader definition of life, or rather livingness, than is colloquially adopted. This is not a novel concept; Trifonov's definition is far broader than biological life, and IIT extends the concept of consciousness to many systems generally regarded as inanimate. Another similar thesis is the concept of 'Lyfe' that contains as one example the familiar organic-biological life of terrestrial ecosystems but is not solely comprised of that. (Bartlett, 2020).

On the basis that many definitions of life are as noted strongly informational (Vogel, 2017), the authors defined a set of broad, abstract biosignature concepts, and then attempted to search for these in the class of simple mathematical systems known as Lifelike Cellular Automata (LCAs). This publication will further elaborate on that approach, correct previous work, and attempt to consider the reverse; *can patterns in the set of LCAs be used to define biosignatures in other living systems*?

## Cellular Automata. Rulesets and Evolving Patterns

A cellular automaton (CA) is a mathematical system that acts as a generalized abstraction of the principles of biological living things. A CA consists of a group of discrete units, referred to as 'cells'. Each cell has a finite number of internal states, and a finite number of neighbors to which it is connected. Each cell evolves in discrete jumps of time, referred to as 'generations', updating its internal state according to an internal ruleset and the internal states of its neighbors. This results in complex collective behavior of the cells.

The basic idea of units that interact in this way can be used as a mathematical abstraction of the interactional principles that underly neural interactions in the brain, flocking behavior of birds, replication of bacteria, or a litany of other biological processes, and they can be used as models thereof. The authors would contend (McCrum & Kee, 2024) that CAs can in fact be seen as an example of livingness in and of themselves, parallel to, but not simply a reduced model of, biological life.

The most famous example of biology-like behavior in CAs is Conway's Game of Life (CGOL), discovered by John Horton Conway in 1970 (Gardner, 1970; Izhikevich, 2015). CGOL is remarkable because despite its very simple ruleset – especially relative to the first CAs, which were far more elaborate in comparison (Von Neumann, 1966), it can display behavior that seems to the qualitative observer exceedingly complex and 'lifelike', hence the name.

Conway's game of life was discovered by Conway during a quest to find in mathematical state-space a 'universal' cellular automata. This CGOL is universal in the sense that it is Turing complete, which means that any computer program that an idealized computer can carry out can be carried out using Conway's Game of Life. Moreover, it is not in general possible to predict the evolution of a configuration in CGOL for the same reason that it is impossible to predict when a computer running a program will get caught in a loop. (Rendell, 2014).

Figure 1 shows what is arguably the most famous pattern in CGOL, the 'glider'. From a quantitative perspective, this serves as a form of information transfer; it is key to the proof of Turing completeness, and its presence was what caused Conway to realize he had found a special ruleset. From a qualitative perspective, the glider's motion across the board evokes the behavior of a small biological lifeform. Conway remarked that had he been able to see the pattern in action at the time he named it, he would have called it 'the ant'. (Conway, 2014).

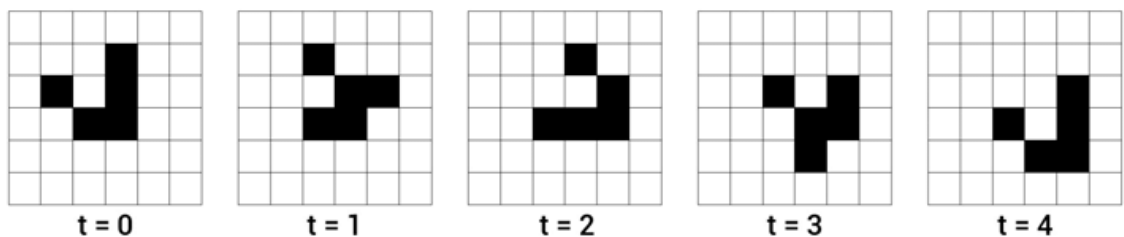

*Figure 1. A CGOL glider in motion. (Dorin, 2012)*

Conway's Game of Life is the namesake and founding example of a broader class of cellular automata, Lifelike Cellular Automata (LCAs). This is the special class of cellular automata with the following traits:

(i) A simple binary state space; each cell can only be *off* or *on*, usually referred to as *live* or *dead*.
(ii) Board consisting of a two-dimensional grid of squares.
(iii) Neighbors to each cell consisting of only of the 8 cells immediately adjacent in the grid- these are referred to as its Moore Neighborhood.
(iv) The state of a cell at generation T + 1 depends only on its state at generation T and its total number of live neighbors.

A particular CA within the LCA class has a given ruleset, defined as the set of responses of live and dead cells to different total numbers of live neighbors. This can be summarized as a pair of binary strings of length 9, meaning that there are $2^{18}$ distinct LCAs. For example, CGOL is the specific LCA whose ruleset is:

$$(0, 0, 1, 1, 0, 0, 0, 0, 0)$$

$$(0, 0, 0, 1, 0, 0, 0, 0, 0)$$

Which should be read as '*a live cell will live if it has two or three live neighbors and die otherwise. A dead cell will become alive if it has three live neighbors and remain dead otherwise*'.

An LCA is defined on a finite grid, which can be toroidal – that is to say, the top left cell is bordered by the bottom right cell to its top left – or boxed – that is to say, the top left cell has no top left neighbor. The specific cells in a grid that are live or dead are referred to as a configuration; these are rotation-symmetric. The percentage of live cells in a given configuration is referred to as its density. The most striking feature of CGOL from a broad perspective across all configurations is a distinct set of phases, dependent on initial configuration density. When the density of a configuration is < 15%, CGOL behaves completely unpredictably. Tiny variations in initial configuration mean the difference between a pattern that explodes in size, a pattern that dies out, or a pattern that stabilises. Much of the computational behaviour and interesting constructions of CGOL are seen in this domain. When the density of a configuration is 15-70%, the initial configuration will display long-lasting and complex evolutions that nonetheless have a strong and predictable attractor at around 2.87%, referred to in the previous publication (McCrum & Kee, 2024) as the Flammenkamp value, after its discoverer. (Flammenkamp, 2004). After a very large number of generations, the vast majority of possible CGOL configurations will evolve towards a density very close to this value. When the density of a configuration is over 70%, the initial configuration will be drawn much more rapidly towards an attractor somewhere lower in density than the Flammenkamp value. For configurations higher than around 85%, this attractor is zero, because highly overpopulated initial configurations will rapidly die out. These phase behaviours are shown in Figures 2 and 3.

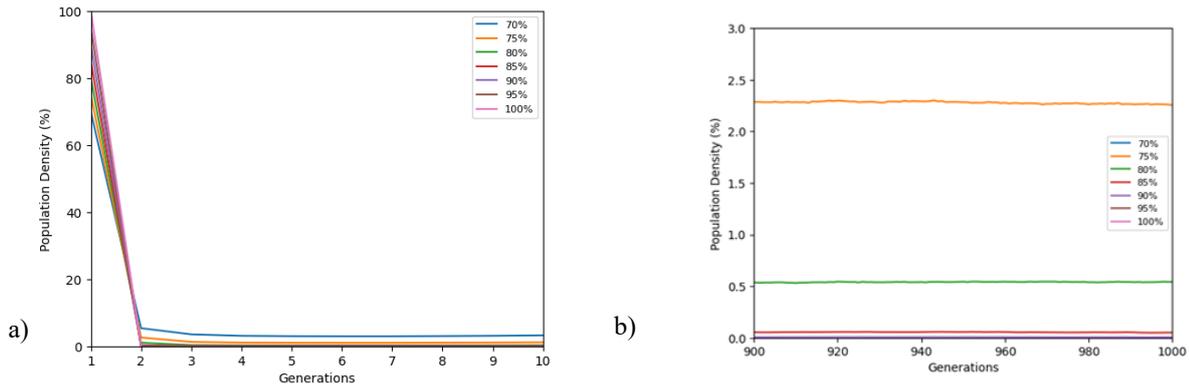

*Figure 2a. The short-term (a) and long-term evolution (b) of random configurations of initial density > 70% in CGOL. As can be seen, the highest density configurations die out nearly instantly; the slightly lower density configurations are rapidly drawn to very small density attractors. This data was produced in Fortran 90 (code available upon request).*

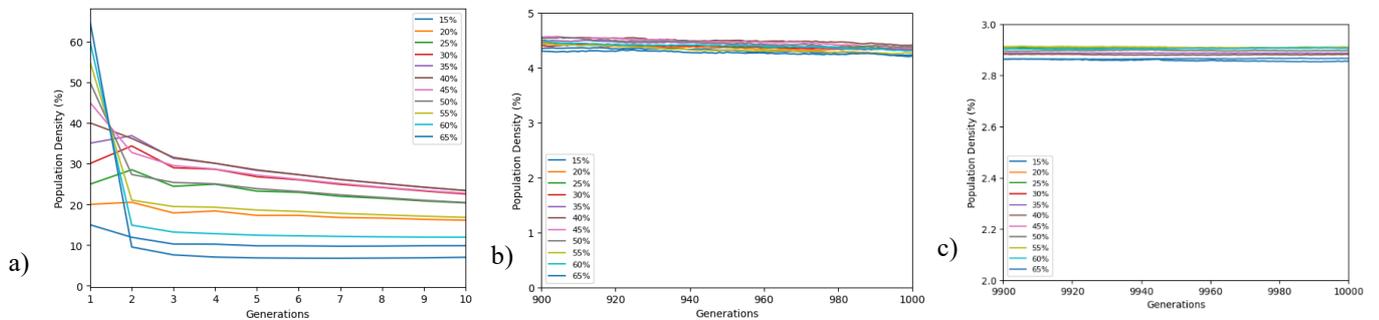

*Figure 3. The short-term (a), long-term (b) and very long-term (c) evolution of random configurations of initial density 15-70% in CGOL. As can be seen, all of the configurations move gradually and continually towards a value of 2.87 – the Flammenkamp Attractor. This data was produced in Fortran 90 (code available upon request).*

Wolfram has produced a fourfold categorisation of 1D and 2D cellular automata rulesets (Wolfram 1984):

(i) CA where almost all patterns die out.
(ii) CA where almost all patterns stabilise into frozen 'still life's or endlessly repeating oscillators.
(iii) CA where almost all patterns explosively grow to fill the board
(iv) CA where a combination of growth and decay produce rich and complex behaviour.

He placed CGOL in the fourth category, arguing that gliders are only possible in this category.

In the previous publication (McCrum & Kee, 2024), the authors argued for three distinctive features of living behaviors that could be identified in CGOL more strongly than other LCAs:

- **Self-Ordering**. Highly living systems are anti-entropic, such that they become more ordered over time in apparent defiance of the Second Law of Thermodynamics. The authors associated this with the Flammenkamp attractor; a wide variety of highly disordered initial states, referred to as 'soups', will inevitably evolve into a specific subset of ordered final states, referred to as 'ashes'. It was shown that while a sample of LCAs often also have ordering attractors, CGOL's is somewhat further from equilibrium than average.

- **Criticality**. Living systems exist near phase transitions, which allows them to maximize information processing and transfer. The authors associated this with the complex behavior of CGOL in the 0-20% density range and Wolfram's classification, arguing that there is an analogy between, for example, the brain existing on the cusp of a phase of explosive electrical signal propagation and exponential electrical signal decay, and CGOL existing somewhere between categories of explosive pattern growth and decay. It was shown that in the 0-20% density range a set of random initial configurations have more varying outcomes in CGOL than on average for other LCAs.

- **Integration**. Living systems have parts that are in strong communication such that to remove any one part is to strongly affect the whole. The authors associated this with an interpretation of the Φ parameter from Integrated Information Theory, as a direct measurement of this systemic feature.

These are all well-established features of biological living systems (Schrödinger, 1944; Mandelbrot, 1982; Havlin, 1995; Christensen, 2005; Losa, 2009; Kurakin, 2011; Pross, 2013; Wallace, 2015; Arsiwalla, 2016; Aguilera, 2018; 2019; Kim, 2019; Popiel, 2020; Phillips, 2020; Khajehabdollahi, 2022; Tian, 2022; Walter, 2022; Ansell, 2024; Niizato, 2024) and it is already known that CGOL displays criticality (Bak, 1989; Bak, 1996; Kayama, 2013; Reia, 2014; Peña, 2021; Akgün, 2024). However, in retrospect, the correspondence stated between these features and features of CGOL may not be as exact as implied.

It's not entirely clear that the criticality of CGOL can be equated exactly to Wolfram's phases of growth and decay. As noted, most of the initial configuration's decay slowly towards the Flammenkamp value (Figure 3), and a newer classification of CAs (Eppstein, 1999) is more exacting about definitions, dividing them into those where no pattern can contract, no pattern can expand, and patterns can both expand and contract.

While both the human brain and CGOL balance growth and decay in a similar fashion, the measurement of variance from initial conditions is not a robust test of criticality (even if CGOL is already known to be critical). Although strong sensitivity to external perturbations is a hallmark of criticality, it is completely possible for a system to be highly sensitive to initial conditions while not being critical, and it is important to be specific in defining criticality. (Beggs, 2012). The wider application of this result should be taken with care, especially as there is some evidence that living systems do not always optimally operate at exactly the point of criticality. (Khajehabdollahi, 2022)

The results in computing Φ were also problematic. Putting aside the criticisms of IIT, a major practical criticism of IIT is that calculating Φ is extremely computationally intensive. The study (McCrum & Kee 2024) was only able to look at Φ for extremely small networks of cells interacting under CGOL rules, and the results were non rotation-symmetric – which seems likely erroneous – and placed CGOL below the average of Φ. CGOL is already known to be a critical system, and there is some evidence that Φ is maximized at the point of criticality in living systems, which may further indicate problems in the calculations. (Niizato, 2024)

Overall, while the previous publication was important in that it established the concept of a biosignature defined using outlier testing of CGOL in an LCA set, it was flawed and limited. This publication looks to build on that basis. Instead of using already established features of living systems and trying to map them to moderately outlier traits of CGOL, it would be more useful to now look at a much more distinctly extreme – relative to other LCAs – feature of CGOL and instead try to derive a possible biosignature concept from it.

## Injectivity, Surjectivity and Informational Memory

In the 1960s, before Conway even discovered CGOL, Moore (Moore, 1962) and Myhill (Myhill, 1963) proved that a broad class of cellular automata – including all LCAs – have equality of two properties: *injectivity* and *surjectivity*. Their proof is referred to as the Garden of Eden theorem.

LCAs are deterministic, so that any given configuration, $c_T$, will always predictably evolve into the same configuration in the succeeding generation, $c_{T+1}$. However, this does not make them necessarily reversible, in that knowing $c_T$ does not inherently tell you $c_{T-1}$. It is possible for two or more configurations to evolve into the same end-state, which means that there is a degeneracy in possible parents for a configuration in general.

To say that an LCA is *injective* is to say that it is completely reversible, such that for any given $c_T$ there is only one possible $c_{T+1}$ and one possible $c_{T-1}$; there is never a degeneracy in possible parents for any possible configuration. To say that an LCA is *surjective* is to say that every configuration has a possible predecessor and therefore can evolve naturally into existence over the course of ordinary play, without external interference.

To demonstrate in real terms what this means, Figure 4 shows a set of evolutions for a trivially injective-surjective LCA – the identity LCA, which maps any input configuration to itself – and for CGOL, which is not injective – as is shown by the fact that the vast majority of input states converge to the relatively small number of configurations that have densities close to the Flammenkamp value – and therefore not surjective.

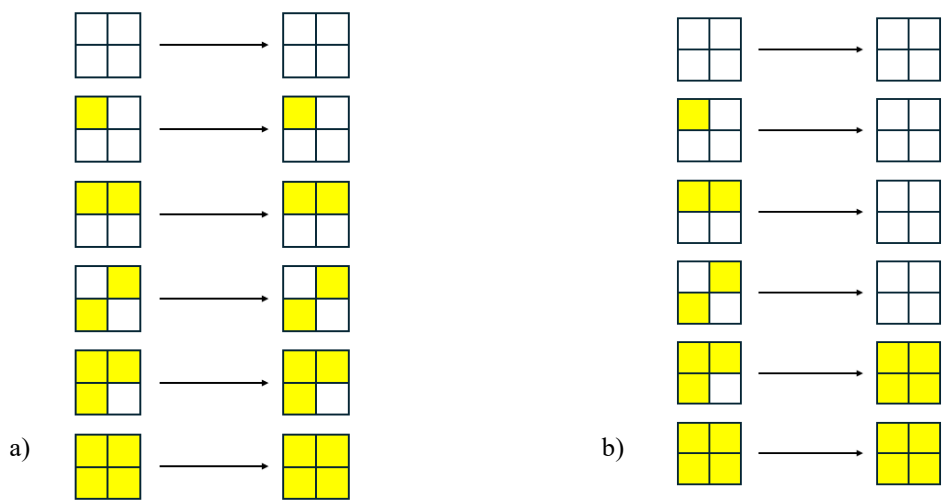

*Figure 4a. The first evolution of every single rotation-distinct possible configuration of a 2x2 box grid under the identity LCA. Note that the system begins with six distinct configurations, and ends with six distinct configurations, and that each of the successor configurations has exactly one predecessor.*

*Figure 4b. The first evolution of every single rotation-distinct possible configuration of a 2x2 box grid under CGOL. Note that the system begins with six distinct configurations and ends with two distinct configurations, and that as a result there are six – two = four configurations that have no possible predecessor; these are referred to as Gardens of Eden.*

It's able to be gathered from Figure 2 why the Garden of Eden theorem is true. Taking the total set of possible configurations of some finite grid $C_0$ and running one iteration of an LCA, an injective LCA will map each configuration to exactly one other, ensuring that each will have a single parent and a single child. A non-injective LCA will map some configurations repeatably onto the same child configuration, leaving others without parents – the Garden of Eden patterns, which do not appear in the second generation and cannot be created through evolution of any possible configuration of the system. They are so called because they can only appear through the outside 'divine' intervention of a player

From a naïve perspective, if one wanted to define the concept of 'memory', or retention of information about the system's past, in an LCA, one could think that the subclass of injective-surjective LCAs has perfect memory. Given a configuration, it is always possible to define exactly which configuration came before, and before that, and so on. This contrasts with CGOL, which is irreversible in a way that destroys information; because the trajectories of separate starting points in configuration space merge, there is no way to tell from the end point which state the system started in.

From an information theory perspective, one could define the information entropy, $H$, (Shannon, 1948) of a configuration in an LCA as being:

$$H = p_1 log\left(\frac{1}{p_1}\right) + p_2 log\left(\frac{1}{p_2}\right) .. + p_n log\left(\frac{1}{p_n}\right)$$

Where $p_n$ is the probability of the nth possible parent configuration being the predecessor of that state. In the case that there is only one possible predecessor:

$$H = 1 log\left(\frac{1}{1}\right) = 0$$

Therefore, any configuration in an injective-surjective LCA has perfect minimisation of information entropy, defined this way. The vast majority of LCAs are not injective-surjective (Figure 5). However, what is important to recognise is that CGOL, despite not having perfect reversibility, has a sense of its past that an injective-surjective LCA cannot possibly have. Once an LCA enters a loop or becomes frozen, it ceases to have a sense of time. There is no way to tell if the system has been progressing for 10 generations or for 10,000 generations. Successive states passing through the system do not deposit any information in the system.

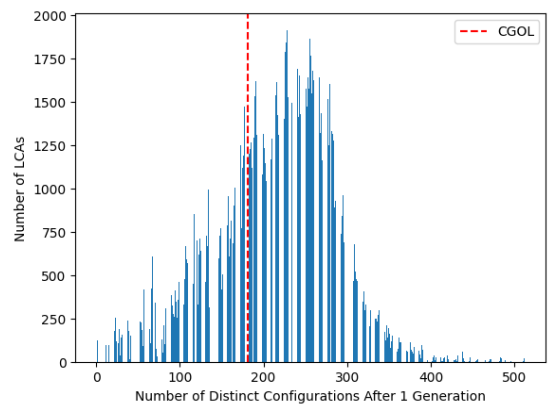

*Figure 5. Number of distinct configurations after one evolution of every single LCA on a 3x3 grid. As a 3x3 grid has 9 cells, it has $2^9 = 512$ possible initial distinct configurations. Only those LCAs with 512 configurations after one evolution are injective-surjective; as can be seen, this is a tiny number at the extremity of a distribution that resembles a skewed bell-curve. CGOL is just to the left of the peak of the curve; it has 181 distinct configurations after one evolution. This data was produced in Fortran 90 (code available upon request).*

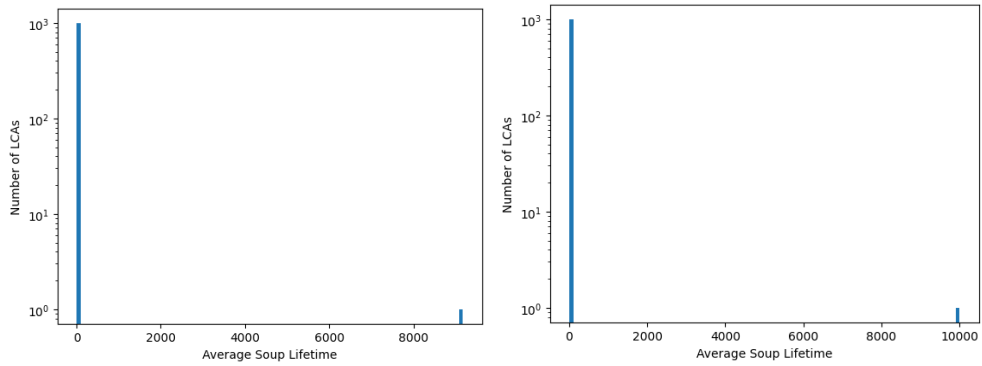

*Figure 6. The average (over 10 samples per ruleset) soup lifetimes (measured until a grid is emptied or enters an oscillation) of 1000 non-CGOL LCA rulesets and CGOL, for a box grid (left) and a torus grid (right), both of size 2048. CGOL's lifespan – the single instance in the right bar on both histograms- is enormously larger than every single other LCA considered – the other 1000 rulesets in the left bar on both histograms. This data was produced in Fortran 90 (code available upon request).*

CGOL is extremely unusual in that for the vast majority of its initial configurations, it continues to have a sense of time for a very large number of generations. This can be seen in Figure 3, which shows a much more extreme outlier behaviour than was identified in the previous study (McCrum & Kee 2024).

The fact that CGOL is not injective-surjective is not a defect but can be considered an aspect of its *biosignature*. An injective-surjective LCA does not preserve evidence of external intervention in Garden of Eden states, nor does it allow for the age of a system to be determined. Because every state has a parent, there is no end or beginning; any given state can feasibly be located within a loop. While some states in CGOL occupy loops in configuration space, others can be traced back to Garden of Eden states, indicating a definitive earliest point where the game began.

Furthermore, the fact that CGOL narrows the configuration space over time produces an "arrow of time" of sorts in the system. The previous publication drew an analogy between this and the evolution of biological systems to more ordered and complex states. Examining a set of CGOL configurations evolving over many generations, the number of distinct configurations will decrease over time, and the density will most likely tend towards the Flammenkamp value; this is clearly asymmetric in time, as is shown in Figure 3.

An injective-surjective LCA will not demonstrate these behaviours, and even other non-injective, non-surjective LCAs, as Figure 4 shows, will very rapidly enter a loop or freeze, and cease to evolve in time. CGOL is unusual because its initial configurations either evolve for a very long time before stopping or never cease to evolve; some configurations studied by the authors were still changing after tens of thousands of generations.

## Temporal Retention of Information as a Biosignature

The previous section has shown that Trifonov's definition of life as *self-replication with variations*, is perfectly captured in the long-term behaviors of Conway's Game of Life.

The term that we will use for this is '*temporal retention of information*'. It's tempting to refer to it as 'memory', but the case of injective-surjective LCAs provide a warning against using that language. Injective-surjective LCAs can be said to have perfect memory, in that configuration $C_T$ 'remembers' that it came from configuration $C_{T-1}$ and previously from configuration $C_{T-2}$, and so on. However, what these systems lack is a sense of time. For example, the identity LCA does have perfect information transfer in the sense that nothing is lost from generation to generation, but no actual meaningful information is being passed on, because the system is not changing. Additional generations do not bring anything new to the system, and the passage of time is not registered. Therefore, we will use the term *temporal retention of information* to convey that such a system does not just have an (albeit imperfect) informational link to its past, but also registers the passing of time through changes in the inter-generational transfer of information.

Any analogy to biological systems should not be treated as a 1:1 correspondence. While the authors would argue that there is no conceptual distinction in terms of the phenomenon of livingness between CGOL and a human brain, a human brain is not evolving towards a continuous fixed point in the same way that CGOL is.

What is important here is that CGOL balances both replication – that is, the perfect passage of information from present to future – and evolution – that is, the temporally asymmetric alteration of information from present to future. In one sense CGOL can be said to rest on a knife-edge between growth and decay, as Wolfram argued. However, in another sense, it rests on a knife-edge between remembering and forgetting, and this is also true of biological systems.

For example, the glider cannot exist in systems that rapidly explode or decay every pattern. Although these systems have short-term strong temporal asymmetry, essentially all initial configurations will end up the same way, and at that point the system ceases to evolve. The glider maintains its form only in systems that allow for both the copying and the mutation of information forwards in time. In one sense, it is the same structure, and in another sense, an entirely different one. This is a philosophical concept referred to by Deleuze as 'difference and repetition' (Deleuze, 1968; Grosz, 2007). The broader idea that a optimal living system's memory should not have perfect fidelity is not novel to this publication. (Levin, 2024) To quote Levin directly:

*There is a paradox which points out that if a species fails to change, it will die out, but if it changes, it likewise ceases to exist.*

Although this may seem like a purely philosophical or even semantic argument, it can be placed in more precise terms by noting that the idea of temporal retention of information is somewhat of a generalization of the concept of integration established by IIT. The parameter of $\Phi$ is, mechanically, a measurement of the responsivity of a system to damage; it measures the divergence in state of a system if part of that system is severed from the rest of the system. Although originally conceived in terms of neural networks, this can be equally thought of as measuring the immune response of the body to a cut, or response of a group of people to an injured member. A system is more integrated when a change in one of its parts produces a response in the system. Nonetheless, this response cannot be overly enhanced; that would correspond, for example, to an overcritical state of the brain where a single firing neuron triggers an electric cascade. For another good example, reproduction by cloning or familial inbreeding ensures perfect fidelity in transmission of a signal but does not allow for suppression of error propagation or evolution.

Purely in terms of the interactions between the cells in a cellular automaton, there is integrated information that is definable using $\Phi$, but temporal retention of information goes beyond this. There is also information stored in, for example, the presence of a Garden of Eden in a configuration's tree of possible histories, or the changing density over time. This is a broader and more complex form of memory, and one that is strongly analogous to that seen in the evolution and growth of biological living things.

One important caveat, previously noted as a strong criticism of IIT, (Nizami, 2019), is that information is *observer dependent.* Although a game of CGOL destroys information as it moves from generation to generation even under ideal circumstances, if there is no knowledge of the ruleset, there is no way whatsoever to reconstruct a parent configuration from a child configuration. The information about previous states is not explicitly stored in a configuration, but in the known ruleset that was used to produce that configuration. For example, the possible parents of a configuration under the identity ruleset and CGOL differ completely.

For an equivalent example, it is possible to use carbon-dating to reconstruct the age of matter taken from a biological organism. However, this information is not freely accessible in the matter itself, but instead it requires existing knowledge of nuclear physics. A person without knowledge of the 'ruleset' of radioactivity decay would not be able to inherently deduce anything about the history of the matter in its current state. Even just a few centuries ago, humans would not have had this information.

Therefore, it is important when discussing temporal retention in an informational sense to ensure that the same amount of information is accessible about all systems under comparison. For example, one can only compare two cellular automata if you know both their rulesets.

## Conclusion

In a previous publication, the authors drew an analogy between CGOL, which is an outlier in the broader set of LCAs, and a world with a biosignature, which is as an outlier in the broader set of abiotic worlds. (McCrum & Kee 2024). If this was only an analogy, then this would be of little use to astrobiologists. It's perfectly possible for two situations to look alike on a surface level but fundamentally have little in common. Two things being outliers in their sets can be a meaningless comparison.

However, the point of that paper was to argue that the *specific outlier traits* of a biologically active world and of CGOL are shared. A set of traits known to be especially characteristic of biological organisms were identified, and their presence in CGOL tested for. The paper had some partial success in showing that CGOL is not just an outlier like biological life is, but it is an outlier in the *same way* that biological life is.

The major objection to considering LCAs is that they are trivial – mere toy models. In one sense, this is completely true; Conway's Game of Life does not remotely approach the complexity of functioning of even the simplest biological organism's brain. In another sense, this misses the point. Because they are 'trivial', they provide an avenue into understanding fundamental principles of life in a way that more complex systems make difficult. To quote Trifonov, discussing his definition of what 'life' is:

*"One unforeseen property of the minimalistic definition is its generality. It can be considered as applicable not just to 'earthly' life but to any forms of life imagination may offer, like extraterrestrial life, alternative chemistry forms, computer models, and abstract forms. It suggests a unique common basis for the variety of lives: all is life that copies itself and changes."*

CGOL and other LCAs are contained by Trifonov's definition, as are many esoteric conceptions of forms or categories of life that predate this paper; to list only a few, astrophysical life (Smolin, 2004; Vidal, 2016; Matloff, 2016; Vidal, 2020; van Duin, 2025), the shadow biosphere (Cleland, 2007; Davies, 2009), the Gaia system (Lovelock, 1972), dynamic kinetic stability (Pross, 2004; 2013; Cutts, 2025), and autocatalytic sets (Hordijk, 2010; 2013). With these previous concepts in mind, the idea of a broader conception of life or livingness (used here and in previous publications by the authors similarly to how the term 'Lyfe' is used by Bartlett and Wong to mean concepts that go beyond the standard biological understanding of the term life (Bartlett, 2020) should not seem overly shocking or problematic.

It may be of value to abandon the implicit assumption that life is a binary, as discussed in another previous publication. (Kee & McCrum, 2024). Viruses present just one example of how a strict living/non-living binarism can be problematic. (Lwoff, 1957; Rybicki, 1990; Forterre, 2010; Koonin, 2012; 2016; de la Higuera, 2022). Furthermore, while CGOL is an outlier – especially with regards to the extreme outlier behaviour shown in this paper – it exists on a continuum, often sharing some of its most distinctive traits with other members of the LCA. This is equally true of biosignatures, of course; they arise as continuous points within a landscape of different features of worlds in the universe, not totally aberrant and separated phenomena unlike anything else seen. Biological systems, philosophically, are said to be *natural kinds* (Papale, 2023) The way that biosignatures are understood is already well-disposed for a spectrum approach.

Ultimately, neuroscience may provide a useful model. As noted, IIT is not necessarily a good model of consciousness in its functional aspects. However, it has been argued that one can use a form of 'weak IIT' (Mediano, 2022), or perhaps an 'IIT inspired' approach (Leung, 2023), that takes inspiration from the philosophy of IIT without necessarily being slavishly adherent to its more criticised aspects. In this way, IIT does not distinguish in binary terms between the conscious and non-conscious but instead defines a quantitative spectrum of the parameter Φ – which 'strong IIT' takes as directly equivalent to consciousness, but weaker approaches may take as only a correlate – that is equally applicable to a computer circuit or to a dog.

Perhaps this is a useful approach for astrobiology to consider adopting. Instead of the binary distinction of life/non-life that causes both severe philosophical difficulties of definition and practical difficulties of detection, one could imagine instead defining a spectrum of livingness in terms of an informational correlate. This would be analogous to Φ and could be used to place different systems on different levels of biosignature, being able to compare like with like using the universal language of information, even if the underlying structures are functionally very different.

Future research may consider looking at the possibility of using measurements such as criticality, Φ (or a modification thereof), or the concept of temporal retention proposed in this study to place different forms of systems on a spectrum. It should be possible to hopefully move from the understanding offered by cellular automata - which are, as noted, a very important case study, but only one singular example - of how different systems can be differentiated informationally to an operational ranking of physical, astronomical systems in terms of degree of livingness. In other words, it should be possible to develop a broader and more comprehensive understanding of what life is without ever having to have left Earth itself.

**Acknowledgements**: The authors would like to thank Gaelan Steele for her help at every stage in this process, as well as Sina Khajehabdollahi, Eve Moore, and Mia Corliss for their helpful insights and commentary. This work received no funding.

# References

Abrahao, F.S.; Santiago Hernández-Orozco, S.;Kiani, N.A.; Tegner, J.; Zenil, H. (2024). Assembly Theory is an approximation to algorithmic complexity based on LZ compression that does not explain selection or evolution. *PLOS Complex Systems*, 1(1): https://doi.org/10.1371/journal.pcsy.0000014.

Aguilera, M. &. D. Paolo E., (2018). Integrated information and autonomy in the thermodynamic limit. *The 2018 Conference on Artificial Life*. pp. 113-120: https://doi.org/10.48550/arXiv.1805.00393

Aguilera, M. (2019) Scaling behaviour and critical phase transitions in integrated information theory *Entropy* 21, 1198; doi:10.3390/e21121198.

Akgün, H., Yan, X., Taşkıran, T., Ibrahimi, M., Mobaraki, A., Lee, C.H. (2024). Deterministic criticality & cluster dynamics hidden in the Game of Life.  10.48550/arXiv.2411.07189.

Albantakis, L., Barbosa, L., Findlay, G., Grasso, M., Haun, A. M., Marshall, W., Mayner, W. G. P., Zaeemzadeh, A., Boly, M., Juel, B. E., Sasai, S., Fujii, K., David, I., Hendren, J., Lang, J. P., & Tononi, G. (2023). Integrated information theory (IIT) 4.0: Formulating the properties of phenomenal existence in physical terms. *PLoS Computational Biology*, *19*(10), e1011465. https://doi.org/10.1371/journal.pcbi.1011465

Ansell, H. S., & Kovács, I. A. (2024). Unveiling universal aspects of the cellular anatomy of the brain. *Communications Physics*, *7*(1). https://doi.org/10.1038/s42005-024-01665-y

Arsiwalla, X. D., & Verschure, P. F. M. J. (2016). High integrated information in complex networks near criticality. In *Lecture notes in computer science* (pp. 184–191). https://doi.org/10.1007/978-3-319-44778-0_22

Bak, P. (1996). The "Game of Life": Complexity is Criticality. In: How Nature Works. Copernicus, New York, NY. https://doi.org/10.1007/978-1-4757-5426-1_6

Bak, P., Chen, K., & Creutz, M. (1989). Self-organized criticality in the 'Game of Life". *Nature*, *342*(6251), 780–782. https://doi.org/10.1038/342780a0

Barrett, A., & Mediano, P.A.M. (2019). The Phi measure of integrated information is not well-defined for general physical systems. *Journal of Consciousness Studies: controversies in science and the humanities*, *26 (*1-2*)*, pp.11-20, https://arxiv.org/abs/1902.04321

Bartlett, S., & Wong, M. L. (2020). Defining Lyfe in the universe: from three privileged functions to four pillars. *Life*, *10*(4), 42. https://doi.org/10.3390/life10040042

Bartlett, S., Eckford, A. W., Egbert, M., Lingam, M., Kolchinsky, A., Frank, A., & Ghoshal, G. (2025). Physics of Life: Exploring information as a distinctive feature of living systems. *PRX Life*, *3*(3). https://doi.org/10.1103/rsx4-8x5f

Beggs, J. M., & Timme, N. (2012). Being critical of criticality in the brain. *Frontiers in Physiology*, *3*, 163. https://doi.org/10.3389/fphys.2012.00163

Benner, S.A. (2010). Defining life. *Astrobiology*, 10(10),1021-30. doi:10.1089/ast.2010.0524

Christensen, K. and Moloney, N.R. (2005). Complexity and Criticality. World Scientific Publishing Company.

Cleland, C.E. (2007). Epistemological issues in the study of microbial life: alternative Terran biospheres? *Stud Hist Philos Biol Biomed Sci,* 38(4), pp.847–861.  https://doi.org/10.1016/j.shpsc.2007.09.007

Conway, J.H. (2014). Does John Conway hate his Game of Life? [online] Available at: https://www.youtube.com/watch?v=E8kUJL04ELA  [Accessed 19th November 2025].

Cornish-Bowden, A. and Cárdenas, M.L. (2020). Contrasting Theories of Life: Historical Context, Current Theories. In search of an ideal theory. *Biosystems*, p.104063. https://doi.org/10.1016/j.biosystems.2019.104063.

Cornish-Bowden, A. (2015). Tibor Gánti and Robert Rosen: Contrasting approaches to the same problem. Journal of Theoretical Biology, 381, 6–10. https://doi.org/10.1016/j.jtbi.2015.05.015

Cowie, C. (2023). New work on biosignatures. *Mind*, *133*(530), 452–471. https://doi.org/10.1093/mind/fzad050

Cutts, E., (2025). Solid, liquid, gas...life?. *New Scientist*, 25 April, pp. 34-37, https://doi.org/10.1016/S0262-4079(25)00671-2.

Davies, P. C., Benner, S. A., Cleland, C. E., Lineweaver, C. H., McKay, C. P., & Wolfe-Simon, F. (2009). Signatures of a shadow biosphere. *Astrobiology*, *9*(2), 241–249. https://doi.org/10.1089/ast.2008.0251

de la Higuera, I. and Lázaro, E. (2022). Viruses in astrobiology. *Frontiers in Microbiology*, 13. https://doi.org/10.3389/fmicb.2022.1032918

Deleuze, G. (2014). Difference and Repetition. Translated by P. Patton. London: Bloomsbury.

Doerig, A., Schurger, A., Hess, K. and Herzog, M.H. (2019). The unfolding argumdent: Why IIT and other causal structure theories cannot explain consciousness. *Consciousness and Cognition*, 72, pp.49–59. https://doi.org/10.1016/j.concog.2019.04.002

Dorin, A., McCabe, J., McCormack, J., Monro, G. and Whitelaw, M. (2012). A framework for understanding generative art. *Digital Creativity*, 23(3-4), pp.239–259. https://doi.org/10.1080/14626268.2012.709940.

Dunér, D. (2019). The history and philosophy of biosignatures. In B. Cavalazzi, & F. Westall (Eds.), Biosignatures for Astrobiology (pp. 303-338). (Advances in Astrobiology and Biogeophysics). Springer. https://doi.org/10.1007/978-3-319-96175-0_15

Eppstein, D. (1999). Gliders and Wolfram's Classification. [online] Uci.edu. Available at: https://ics.uci.edu/~eppstein/ca/wolfram.html [Accessed 19 Nov. 2025].

Flammenkamp, A. (2004). Most seen natural occuring ash objects in Game of Life. [online] Uni-bielefeld.de. Available at: https://wwwhomes.uni-bielefeld.de/achim/freq_top_life.html [Accessed 19 Nov. 2025].

Fleming, S.M., Frith, C.D., Goodale, M.A., Lau, H., LeDoux, J.E., Lee, A., Michel, M., Owen, A.M., Megan and Slagter, H.A. (2023). The Integrated Information Theory of Consciousness as Pseudoscience. https://doi.org/10.31234/osf.io/zsr78

Forterre, P. (2010). Defining Life: The Virus Viewpoint. *Origins of Life and Evolution of Biospheres*, *40*(2), 151–160. https://doi.org/10.1007/s11084-010-9194-1

Gardner, M. (1970). Mathematical Games: The fantastic combinations of John Conway's new solitaire game "life." Scientific American, 223(4), 120–123.

Gillen, C., Jeancolas, C., McMahon, S., & Vickers, P. (2023). The call for a new definition of biosignature. *Astrobiology*, *23*(11), 1228–1237. https://doi.org/10.1089/ast.2023.0010

Grosz, E. (2007). Deleuze, Bergson and the concept of life. *Revue Internationale De Philosophie*, *n° 241*(3), 287–300. https://doi.org/10.3917/rip.241.0287

Havlin, S., Buldyrev, S.V., Goldberger, A.L., Mantegna, R.N., Ossadnik, S.M., Peng, C.-K. ., Simons, M. and Stanley, H.E. (1995). Fractals in biology and medicine. *Chaos, Solitons & Fractals*, 6, pp.171–201. https://doi.org/10.1016/0960-0779(95)80025-c

Helman, D. S. (2022). Finding or creating a living organism? past and future thought experiments in astrobiology applied to artificial intelligence. *Acta Biotheoretica*, *70*(2), 13. https://doi.org/10.1007/s10441-022-09438-2

Herzog, M. H., Schurger, A., Doerig, A., Herzog, M. H., Schurger, A., & Doerig, A. (2022). First-person experience cannot rescue causal structure theories from the unfolding argument. *Consciousness and Cognition*, *98*, 103261. https://doi.org/10.1016/j.concog.2021.103261

Hordijk, W. (2013). Autocatalytic Sets: From the Origin of Life to the Economy, *BioScience*, 63(11), 877-881 https://doi.org/10.1525/bio.2013.63.11.6

Hordijk, W., Hein, J. and Steel, M. (2010). Autocatalytic Sets and the Origin of Life. *Entropy*, 12(7), pp.1733–1742. https://doi.org/10.3390/e12071733

Klincewicz, M., Cheng, T., Schmitz, M., Sebastián, M. Á., Snyder, J. S., Arnold, D. H., Baxter, M. G., Bekinschtein, T. A., Bengio, Y., Bisley, J. W., Browning, J., Buonomano, D., Carmel, D., Carrasco, M., Carruthers, P., Carter, O., Chang, D. H. F., Charest, I., Cherkaoui, M., . . . Wheatley, T. (2025). What makes a theory of consciousness unscientific? *Nature Neuroscience*, *28*(4), 689–693. https://doi.org/10.1038/s41593-025-01881-x

Izhikevich, E., Conway, J. and Seth, A. (2015). Game of Life. Scholarpedia, 10(6), p.1816. https://doi.org/10.4249/scholarpedia.1816

Yoshihiko Kayama (2013). Network representation of the game of life and self-organized criticality. https://doi.org/10.1109/alife.2013.6602432

Kee, T. & McCrum, J. (2024). Re-evaluating Boundary Conditions of the Concept of Life. *IAU Proceedings* 387. https://philsci-archive.pitt.edu/27105/

Khajehabdollahi, S., Prosi, J., Giannakakis, E., Martius, G., & Levina, A. (2022). When to be critical? performance and evolvability in different regimes of neural ising agents. *Artificial Life*, *28*(4), 458–478. https://doi.org/10.1162/artl_a_00383

Kim, H., & Lee, U. (2019). Criticality as a determinant of integrated information Φ in human brain networks. *Entropy*, *21*(10), 981. https://doi.org/10.3390/e21100981

Koonin, E.V. (2012). Defining Life: An Exercise in Semantics or A Route to Biological Insights? *Journal of Biomolecular Structure and Dynamics*, 29(4), pp.603–605. https://doi.org/10.1080/073911012010525000

Koonin, E. V., & Starokadomskyy, P. (2016). Are viruses alive? The replicator paradigm sheds decisive light on an old but misguided question. *Studies in History and Philosophy of Science Part C Studies in History and Philosophy of Biological and Biomedical Sciences*, *59*, 125–134. https://doi.org/10.1016/j.shpsc.2016.02.016

Kurakin, A. (2011). The self-organizing fractal theory as a universal discovery method: the phenomenon of life. *Theoretical Biology and Medical Modelling*, *8*(1), 4. https://doi.org/10.1186/1742-4682-8-4

Letelier JC, Cárdenas ML, Cornish-Bowden A. (2011). From L'Homme Machine to metabolic closure: steps towards understanding life. *Journal of Theoretical Biology*, 286(1):100-13. Doi:10.1016/j.jtbi.2011.06.033.

Leung, A. and Tsuchiya, N. (2023). Separating weak integrated information theory into inspired and aspirational approaches. *Neuroscience of Consciousness*, 2023(1). https://doi.org/10.1093/nc/niad012

Levin, M. (2024). Self-Improvising Memory: A perspective on memories as agential, dynamically reinterpreting cognitive glue. *Entropy*, *26*(6), 481. https://doi.org/10.3390/e26060481

Losa, G. A. (2009). The fractal geometry of life. *PubMed*, *102*(1), 29-59. https://pubmed.ncbi.nlm.nih.gov/19718622

Lovelock, J. (1972). Gaia as seen through the atmosphere. *Atmospheric Environment (1967)*, *6*(8), 579–580. https://doi.org/10.1016/0004-6981(72)90076-5

Lwoff, A. (1957). The concept of virus. *Microbiology*, *17*(2), 239–253. https://doi.org/10.1099/00221287-17-2-239

Benoît Mandelbrot (1982). The fractal geometry of nature. New York: W.H. Freeman And Company.

Matloff, G. L. (2016). Can panpsychism become an observational science? *Journal of Consciousness Exploration & Research*, *7*(7). https://jcer.com/index.php/jcj/article/download/579/595

McCrum, J. & Kee, T. (2024). Conways game of life as an analogue to a habitable world Livingness beyond the biological. *IAU Proceedings* 387. https://arxiv.org/abs/2410.22389

Mediano, P.A.M., Rosas, F.E., Bor, D., Seth, A.K. and Barrett, A.B. (2022). The strength of weak integrated information theory. *Trends in Cognitive Sciences* 26(8), pp.646–655. https://doi.org/10.1016/j.tics.2022.04.008

Merker, B., Williford, K., & Rudrauf, D. (2021). The integrated information theory of consciousness: A case of mistaken identity. *Behavioral and Brain Sciences*, *45*, e41. https://doi.org/10.1017/s0140525x21000881

Miller, W.B., František Baluška and Reber, A.S. (2023). A revised central dogma for the 21st century: All biology is cognitive information processing. *Progress in Biophysics and Molecular Biology* 182, pp.34–48. https://doi.org/10.1016/j.pbiomolbio.2023.05.005

Moore, E., 1962. Machine models of self-reproduction. *Proceedings of symposia in applied mathematics*, 14. Doi: 10.1090/psapm/014/9961.

Myhill, J. (1963). Shorter note: the converse of Moore's Garden-of-Eden theorem. *Proceedings of the American Mathematical Society*, *14*(4), 685. https://doi.org/10.2307/2034301

Niizato, T., Sakamoto, K., Mototake, Y., Murakami, H., & Tomaru, T. (2024). Information structure of heterogeneous criticality in a fish school. *Scientific Reports*, *14*(1), 29758. https://doi.org/10.1038/s41598-024-79232-2

Nizami, L., 2019. Information Theory Is Abused In Neuroscience. *Cybernetics and Human Knowing*, 26(4), pp. 47-97. https://philarchive.org/rec/NIZITI

Ozelim, Luan & Uthamacumaran, Abicumaran & Abrahão, Felipe & Hernandez-Orozco, Santiago & Kiani, Narsis & Tegner, Jesper & Zenil, Hector. (2024). Assembly Theory Reduced to Shannon Entropy and Rendered Redundant by Naive Statistical Algorithms. https://arxiv.org/abs/2408.15108

Papale, F. and Montminy, D. (2023). Natural Kinds: The Expendables. *Canadian Journal of Philosophy,* 53(2), pp.103–120. https://doi.org/10.1017/can.2023.30

Peña, E. and Sayama, H. (2021). Life Worth Mentioning: Complexity in Life-Like Cellular Automata. *Artificial Life*, 27(2), pp.105–112. https://doi.org/10.1162/artl_a_00348

Phillips, R. (2020). Schrodinger's "What is Life" At 75: The Physical Aspects of the Living Cell Revisited. *Biophysical Journal*, 118(3), 2a. https://doi.org/10.1016/j.bpj.2019.11.204

Popiel, N.J.M., Khajehabdollahi, S., Abeyasinghe, P.M., Riganello, F., Nichols, E.S., Owen, A.M. and Soddu, A. (2020). The Emergence of Integrated Information, Complexity, and 'Consciousness' at Criticality. *Entropy*, 22(3), pp.339–339. https://doi.org/10.3390/e22030339

Pross, A. (2004). Causation and the origin of life. metabolism or replication first? *Origins of Life and Evolution of Biospheres*, *34*(3), 307–321. https://doi.org/10.1023/b:orig.0000016446.51012.bc

Pross, A. (2013). The evolutionary origin of biological function and complexity. *Journal of Molecular Evolution*, *76*(4), 185–191. https://doi.org/10.1007/s00239-013-9556-1

Reia, S. M., & Kinouchi, O. (2014). Conway's game of life is a near-critical metastable state in the multiverse of cellular automata. *Physical Review E*, *89*(5), 052123. https://doi.org/10.1103/physreve.89.052123

Rendell, P. (2014). Turing Machine Universality of the Game of Life. PhD Thesis. https://uwe-repository.worktribe.com/preview/822581/thesis.pdf

Rybicki, E. (1990). The classification of organisms at the edge of life or problems with virus systematics. *South African Journal of Science*, *86*(4), 182–186. https://journals.co.za/content/sajsci/86/4/AJA00382353_6229

Schrödinger, E., 1944. What is Life: The Physical Aspect of the Living Cell. Cambridge, United Kingdom: Cambridge University Press.

Shannon, C.E. and Weaver, W. (1948). The mathematical theory of communication. Urbana: University of Illinois Press.

Sharma, A., Czégel, D., Lachmann, M., Kempes, C. P., Walker, S. I., & Cronin, L. (2023). Assembly theory explains and quantifies selection and evolution. *Nature*, *622*(7982), 321–328. https://doi.org/10.1038/s41586-023-06600-9

Smolin, L. (2004). Cosmological natural selection as the explanation for the complexity of the universe. *Physica a Statistical Mechanics and Its Applications*, *340*(4), 705–713. https://doi.org/10.1016/j.physa.2004.05.021

Tian, Y., Tan, Z., Hou, H., Li, G., Cheng, A., Qiu, Y., Weng, K., Chen, C., & Sun, P. (2022). Theoretical foundations of studying criticality in the brain. *Network Neuroscience*, *6*(4), 1148–1185. https://doi.org/10.1162/netn_a_00269

Tononi, G. (2004). An information integration theory of consciousness. *BMC Neuroscience*, *5*(1), 42. https://doi.org/10.1186/1471-2202-5-42

Trifonov, E. N. (2011). Vocabulary of definitions of life suggests a definition. *Journal of Biomolecular Structure and Dynamics*, *29*(2), 259–266. https://doi.org/10.1080/073911011010524992

Uthamacumaran, A., Abrahão, F.S., Kiani, N.A. and Zenil, H. (2024). On the salient limitations of the methods of assembly theory and their classification of molecular biosignatures. *npj Systems Biology and Applications*, 10(1). https://doi.org/10.1038/s41540-024-00403-y

Van Duin, M. (2025). Development of Mass and Energy Rate (Density) of Dissipative Systems Over Their Lifetimes: A Comparison of a Low-Mass Star, like Our Sun, a Human, and the Roman Empire. In *World-systems evolution and global futures* (pp. 345–385). https://doi.org/10.1007/978-3-031-85410-1_11

Vidal, C. (2016). Stellivore extraterrestrials? Binary stars as living systems. *Acta Astronautica*, *128*, 251–256. https://doi.org/10.1016/j.actaastro.2016.06.038

Vidal, C. (2020) Energy rate density as a technosignature: The case for stellivores, presentation at Life in the Universe: Big History, SETI and the Future of Humankind: IBHA & INAF-IASF MI Symposium 15–16 July 2019, [Online], http://www.clemvidal.com/s/Vidal-2020-ERD-as-a-technosignature-case-forstellivores.pdf.

Vogel, M. (2017). From matter to life: information and causality, edited by S. I. Walker, P. C. W. Davies and G. F. R. Ellis. *Contemporary Physics*, 58(3), 300–301. https://doi.org/10.1080/00107514.2017.1344307

Von Neumann, J. and Burks, A.W. (1966). Theory of Self-Reproducing Automata. Urbana ; London: Illinois University Press.

Walker, S. I., Davies, P. C. W., & Ellis, G. F. R. (2017). From matter to life: Information and causality. Cambridge University Press.

Wallace R. (2015). Poised for survival: criticality, natural selection, and excitation-transcription coupling. *The international journal of biochemistry & cell biology*, *61*, 1–7. https://doi.org/10.1016/j.biocel.2015.01.013

Walter, N. and Hinterberger, T. (2022). Self-organized criticality as a framework for consciousness: A review study. *Frontiers in Psychology*, 13. https://doi.org/10.3389/fpsyg.2022.911620

Wolfram, S. (1984). Universality and complexity in cellular automata. *Physica D: Nonlinear Phenomena,* 10(1-2), pp.1–35. https://doi.org/10.1016/0167-2789(84)90245-8